\documentclass[iop]{emulateapj}

\usepackage{graphicx}
\usepackage{graphics}
\usepackage{amsmath}
\usepackage{txfonts}
\usepackage{epsfig}
\usepackage{natbib}
\usepackage{times}
\usepackage{amssymb}
\usepackage{color}
\usepackage{longtable}

\newcommand\ergcms{erg\,cm$^{-2}$\,s$^{-1}$}

\newcommand\cmsq{cm$^{-2}$}
\newcommand\integ{{\it{INTEGRAL}}}
\newcommand\swift{{\it{Swift}}}
\newcommand\xmm{{\it{XMM-Newton}}}
\newcommand\chan{{\it{Chandra}}}
\newcommand\nustar{{\it{NuSTAR}}}

\newcommand\suz{{\it{Suzaku}}}
\newcommand\asca{{\it{ASCA}}}
\newcommand\rxte{{\it{RXTE}}}
\newcommand\igr{\object{IGR~J16393$-$4643}}
\newcommand\nh{$N_\mathrm{H}$}

\shorttitle{NuSTAR discovery of a cyclotron line in the accreting X-ray pulsar IGR J16393$-$4643}
\shortauthors{Bodaghee et al.}

\begin{document}

\title{NuSTAR discovery of a cyclotron line in the accreting X-ray pulsar IGR J16393$-$4643}

\author{Arash Bodaghee\altaffilmark{1}}
\author{John A. Tomsick\altaffilmark{2}}
\author{Francesca M. Fornasini\altaffilmark{2,3}}
\author{Roman Krivonos\altaffilmark{4}}
\author{Daniel Stern\altaffilmark{5}}
\author{Kaya Mori\altaffilmark{6}}
\author{Farid Rahoui\altaffilmark{7,8}}
\author{Steven E. Boggs\altaffilmark{2}}
\author{Finn E. Christensen\altaffilmark{9}}
\author{William W. Craig\altaffilmark{2,10}}
\author{Charles J. Hailey\altaffilmark{6}}
\author{Fiona A. Harrison\altaffilmark{11}}
\author{William W. Zhang\altaffilmark{12}}

\altaffiltext{1}{Georgia College and State University, Milledgeville, GA 31061, USA}
\altaffiltext{2}{Space Sciences Laboratory, 7 Gauss Way, University of California, Berkeley, CA 94720, USA}
\altaffiltext{3}{Astronomy Department, University of California, Berkeley, CA 94720, USA}
\altaffiltext{4}{Space Research Institute, Russian Academy of Sciences, Profsoyuznaya 84/32, 117997 Moscow, Russia}
\altaffiltext{5}{Jet Propulsion Laboratory, California Institute of Technology, Pasadena, CA 91109, USA}
\altaffiltext{6}{Columbia Astrophysics Laboratory, Columbia University, New York, NY 10027, USA}
\altaffiltext{7}{European Southern Observatory, Karl-Schwarzschild-Strasse 2, 85748 Garching bei M\"{u}nchen, Germany}
\altaffiltext{8}{Dept. of Astronomy, Harvard University, 60 Garden Street, Cambridge, MA 02138, USA}
\altaffiltext{9}{DTU Space, National Space Institute, Technical University of Denmark, Elektrovej 327, DK-2800 Lyngby, Denmark}
\altaffiltext{10}{Lawrence Livermore National Laboratory, Livermore, CA 94550, USA}
\altaffiltext{11}{Cahill Center for Astronomy and Astrophysics, California Institute of Technology, Pasadena, CA 91125, USA}
\altaffiltext{12}{NASA Goddard Space Flight Center, Greenbelt, MD 20771, USA}

\begin{abstract}
The high-mass X-ray binary and accreting X-ray pulsar \igr\ was observed by \nustar\ in the 3--79 keV energy band for a net exposure time of 50 ks. We present the results of this observation which enabled the discovery of a cyclotron resonant scattering feature with a centroid energy of $29.3_{-1.3}^{+1.1}$ keV. This allowed us to measure the magnetic field strength of the neutron star for the first time: $B = (2.5\pm0.1)\times10^{12}$ G. The known pulsation period is now observed at 904.0$\pm$0.1 s. Since 2006, the neutron star has undergone a long-term spin-up trend at a rate of $\dot P = -2\times10^{-8}$\,s\,s$^{-1}$ ($-$0.6\,s per year, or a frequency derivative of $\dot \nu = 3\times10^{-14}$ Hz\,s$^{-1}$). In the power density spectrum, a break appears at the pulse frequency which separates the zero slope at low frequency from the steeper slope at high frequency. This addition of angular momentum to the neutron star could be due to the accretion of a quasi-spherical wind, or it could be caused by the transient appearance of a prograde accretion disk that is nearly in corotation with the neutron star whose magnetospheric radius is around $2\times10^{8}$\,cm.
\end{abstract}



\section{Introduction}

During a survey of the Galactic Plane, the \asca\ space telescope detected a new X-ray source, \object{AX~J163904$-$4642}, in the direction of the Norma spiral arm tangent \citep{sug01}. The source was initially classified as a microquasar \citep{com04} given the absorbed power-law shape of the \asca-derived spectrum, and given that its X-ray position coincides with counterpart candidates from the radio and infrared bands, as well as with the unidentified gamma-ray source \object{3EG~J1639$-$4702} \citep{har99}. Surveying the same region a few years later, the \integ\ space telescope detected \igr\ which was shown to be the hard X-ray counterpart to the \asca\ source \citep{bir04,mal04}.

A follow-up observation of \igr\ with \xmm\ revealed several interesting clues about the source's nature \citep{bod06}. First, the refined position from \xmm\ excluded all previous multi-wavelength counterpart candidates (other than the \asca\ source) that were the basis for the microquasar interpretation, and pointed instead to a faint object from the Two Microns All-Sky Survey \citep{cut03} named \object{2MASS~J16390535$-$4642137}. Thus, the association of \igr\ with \object{3EG~J1639$-$4702} was attributed to a chance alignment of unrelated objects. Second, the spectral properties were typical of a wind-accreting pulsar \citep[e.g.,][]{nag89}; i.e., a large absorbing column, a hard X-ray continuum with an exponential cutoff around 20\,keV, and iron fluorescence lines. Third, a timing analysis of the \xmm\ and \integ\ data led to the discovery of a coherent pulsation period of 911\,s indicative of a slowly rotating, magnetized neutron star. This pulsation period was confirmed in observations taken with \rxte\ \citep{tho06}, \chan\ \citep{for14}, and \suz\ \citep{isl15}. 

These results suggest that \igr\ is an obscured high-mass X-ray binary (HMXB) in which a compact object (in this case, a spinning neutron star) accretes the wind shed by a massive donor star \citep{whi83,nag89,bil97}. This view is supported by the subsequent detection of a 4.2-d orbital period in \rxte\ and \swift\ data \citep{cor10,cor13,col15} which places \igr\ within the wind-fed HMXB systems in the pulse-vs.-orbital period diagram of \citet{cor86}.

\begin{figure*}[!t]
\begin{center}
\includegraphics[width=\textwidth]{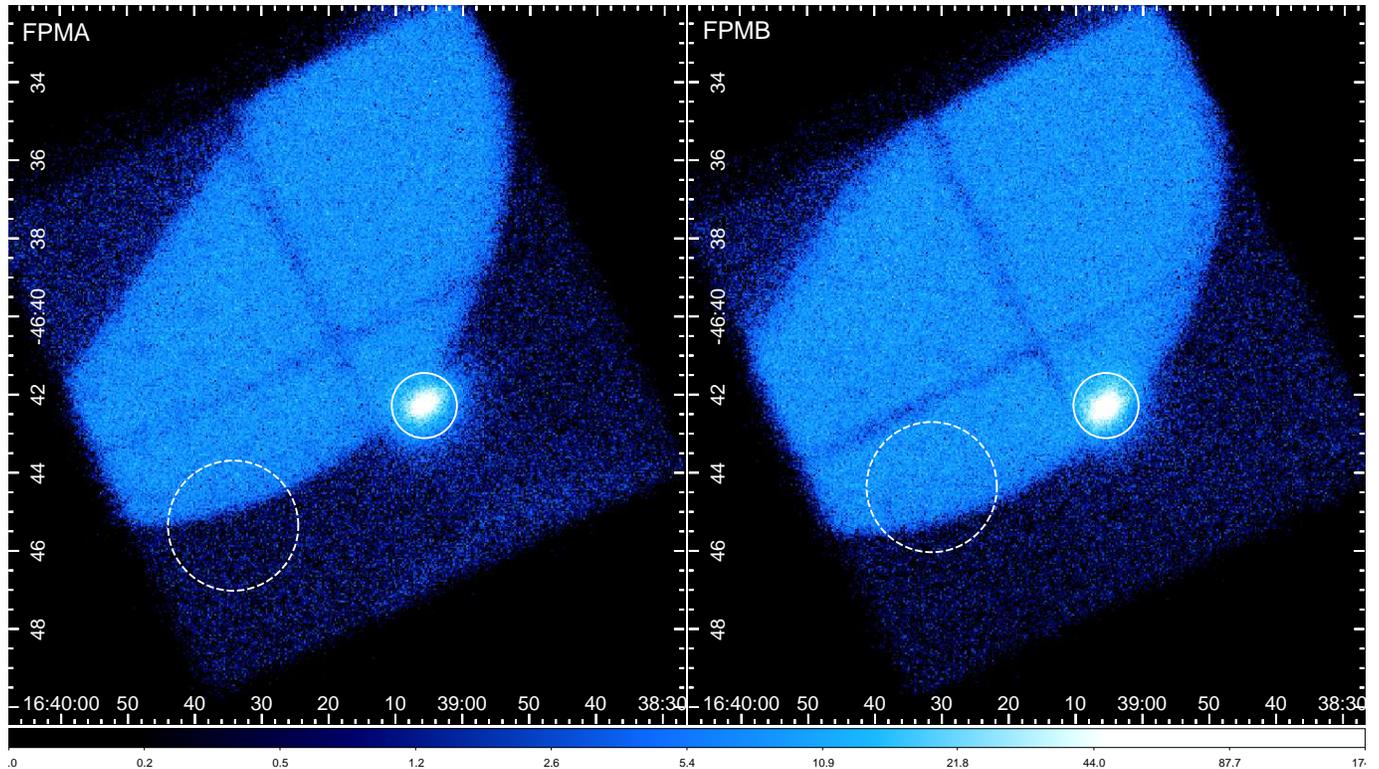}
\end{center}
\caption{Images of IGR~J16393$-$4643 gathered with \nustar\ FPMA (\emph{left}) and FPMB (\emph{right}) in the 3--79\,keV energy band. The images are presented in J2000.0 equatorial coordinates, they are scaled logarithmically, and extraction regions for the source (50$^{\prime\prime}$-radius) and background (100$^{\prime\prime}$-radius) are indicated.}
\label{fig_img}
\end{figure*}

\begin{deluxetable*}{ l c c c c c c }
\tablewidth{0pt}
\tabletypesize{\scriptsize}
\tablecaption{Journal of observations of IGR~J16393$-$4643.}
\tablehead{
\colhead{telescope} & \colhead{observation ID} & \colhead{pointing R.A. (J2000)} & \colhead{pointing decl. (J2000)}	& \colhead{start date (UTC)} 	& \colhead{end date (UTC)} 	& \colhead{effective exposure (ks)} }
\startdata

\nustar\ 		& 30001008002	& 249.8301 & $-$46.6567 		& 2014-06-26 02:21:07 	& 2014-06-27 05:31:07 & 50.579	\\

\swift\		& 00080170004	& 249.6784 & $-$46.6450		& 2014-06-27 04:40:07 	& 2014-06-27 05:27:01 & 2.804 

\enddata
\label{tab_log}
\end{deluxetable*}

However, the exact spectral class of this donor star is uncertain. The optical/infrared spectrum of \object{2MASS~J16390535$-$4642137} indicates a spectral class of BIV-V \citep{cha08}, whereas analysis of the $K_s$-band spectrum of the same object suggests a late-type KM star in a symbiotic binary system \citep{nes10}. To help clarify this issue, we performed a \chan\ observation of \igr\ that provided an X-ray position with sub-arcsecond accuracy \citep{bod12b}. The \chan\ position is R.A. (J2000) $=$ $16^{\mathrm{h}}$ $39^{\mathrm{m}}$ $05\overset{\mathrm{s}}{.}47$ and Decl. $=$ $-46^{\circ}$ $42^{\prime}$ $13\overset{\prime\prime}{.}0$ with an error radius of $0\overset{\prime\prime}{.}6$ (90\% confidence). This error circle excludes the 2MASS star and suggests a faint, blended, and likely distant B star that appears in the mid-infrared at wavelengths longer than 5\,$\mu$m. Indeed, an orbital-period analysis suggests that the donor star is a B giant with a mass greater than 7 $M_{\odot}$ \citep{col15}.

In 2014 June, the \emph{Nuclear Spectroscope Telescope Array} \citep[\emph{NuSTAR}:][]{har13} observed the field of \igr\ as part of its survey of the Norma Arm region \citep{bod14,for14}. \nustar\ provides exceptional angular ($18^{\prime\prime}$ full-width-half-maximum, 58$^{\prime\prime}$ half-power diameter) and spectral resolution (400\,eV) around 10\,keV. In this paper, we present results from these \nustar\ observations of \igr\ in addition to a simultaneous snapshot observation taken with \swift. Section\,\ref{sec_obs} describes the analysis of the X-ray data with results from timing and spectral analyses shown in Section\,\ref{sec_res}. Insights into the nature of \igr\ are discussed in Section\,\ref{sec_disc}.

\section{Observations \& Data Analysis}
\label{sec_obs}

Table\,\ref{tab_log} lists details of the observations of \igr\ included in this analysis. All data were analyzed using HEASoft 6.16. The \nustar\ data consist of the two focal plane modules A and B (FPMA and FPMB) where each module has a field-of-view (FoV) of $13^{\prime} \times 13^{\prime}$. Raw event lists from observation ID (ObsID) 30001008002 were reprocessed with \texttt{nupipeline}, which is part of the \nustar\ Data Analysis Software\footnote{http://heasarc.gsfc.nasa.gov/docs/nustar/analysis/nustardas\_swguide\_v1.7.pdf} (NuSTARDAS 1.4.1), while employing the most recent calibration database files available at the time (CALDB: 2014 August 14). 

Figure\,\ref{fig_img} presents the cleaned images for each module in the 3--79\,keV energy band. From the cleaned event lists of each module, we extracted source spectra and light curves using a 50$^{\prime\prime}$-radius circle centered on the \chan\ position for \igr. Around 70\% of the source photon energy is enclosed within this radius \citep{mad15}. Given the size of this extraction region, and given that the \chan\ position is within 3$^{\prime\prime}$ of the brightest pixel in the image from each \nustar\ focal plane module, we chose not to correct these images for the known systematic offset of coordinates.

The bright background feature affecting both modules (Figure\,\ref{fig_img}) is due to unfocused stray-light photons from \object{GX~340$+$0}, an unrelated object situated just outside the FoV. Since the source extraction regions have some fraction of their area contaminated by stray-light photons, we selected background extraction regions (100$^{\prime\prime}$-radius circles) that encompassed a similar fraction of stray-light photons. Exposure differences due to vignetting were accounted for in the response matrices and spectra. The effective exposure time at the position of \igr\ is 50.579\,ks. 

During the \nustar\ observation, \swift-XRT \citep{bur05} also observed the source yielding an effective exposure time of 2.804\,ks (ObsID 00080170004). We extracted a spectrum in the 0.5--10\,keV energy range that extends the source continuum below the 3-keV limit of \nustar\ thereby enabling the column density to be constrained. The combined \swift-\nustar\ spectra were fit in \texttt{Xspec} 12.8.2 \citep{arn96} where we assumed \citet{wil00} abundances and \citet{ver96} photo-ionization cross-sections.

\begin{figure}[!t]
\begin{center}
\includegraphics[width=0.45\textwidth]{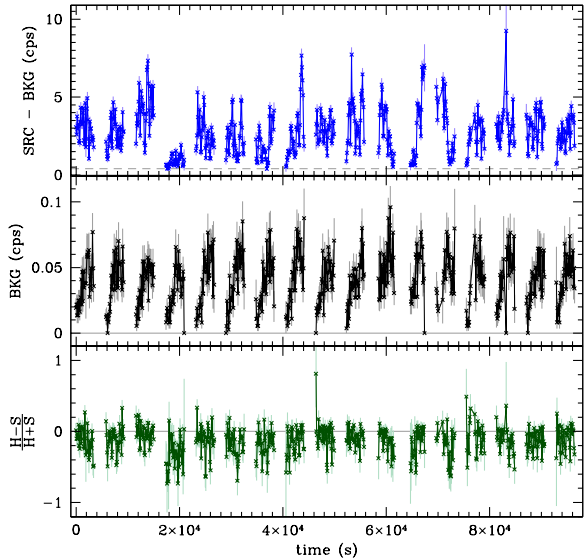}
\end{center}
\caption{Source and background light curves from \nustar\ (3--79\,keV) for IGR J16393$-$4643 where each bin lasts 100\,s. The top panel presents the source light curve combining count rates from FPMA and FPMB that are then background-subtracted. In the middle panel, the summed (FPMA and FPMB) background count rate is shown scaled to the size of the source region. For comparison, the average background rate ($\times10$) is indicated by the dashed line in the top panel. The bottom panel features the hardness ratio where $S$ and $H$ represent count rates in 3--10\,keV and 10--79\,keV, respectively. \vspace{2mm}}
\label{fig_lc}
\end{figure}

\section{Results}
\label{sec_res}

\subsection{Timing Analysis}
\label{sec_time}

Figure \ref{fig_lc} presents the source and background light curves binned at 100\,s in the 3--79\,keV energy range combining counts from \nustar\ FPMA and FPMB. The hardness ratio is defined as $\frac{H - S}{H+S}$ where $S$ and $H$ represent count rates in 3--10\,keV (``soft'') and 10--79\,keV (``hard''), respectively. A dividing value of 10\,keV allocates a roughly even number of source counts between soft and hard energy bands.

\begin{figure}[!t] 
\centering
\includegraphics[width=0.45\textwidth]{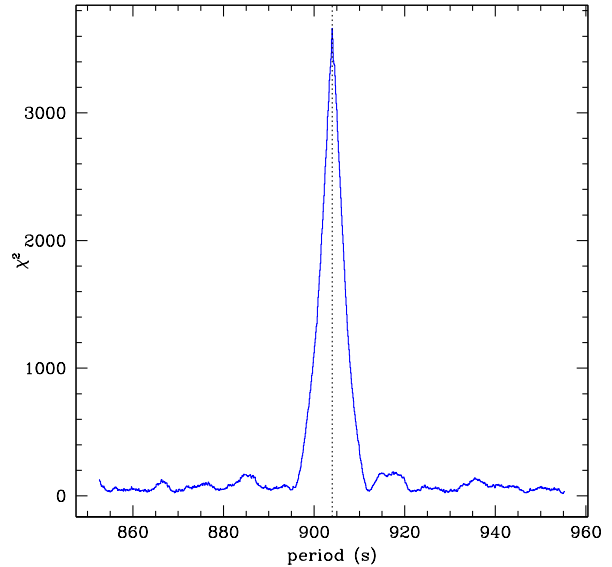}
\caption{Periodicity search ($\chi^{2}$ distribution) on the \nustar\ light curve (3--79 keV) of IGR~J16393$-$4643 centered at 904 s (vertical line), with 20 bins per pulse period, and a resolution of 0.1 s. }
\label{fig_chisq}
\end{figure}

\begin{figure}[!t] 
\centering
\includegraphics[width=0.45\textwidth]{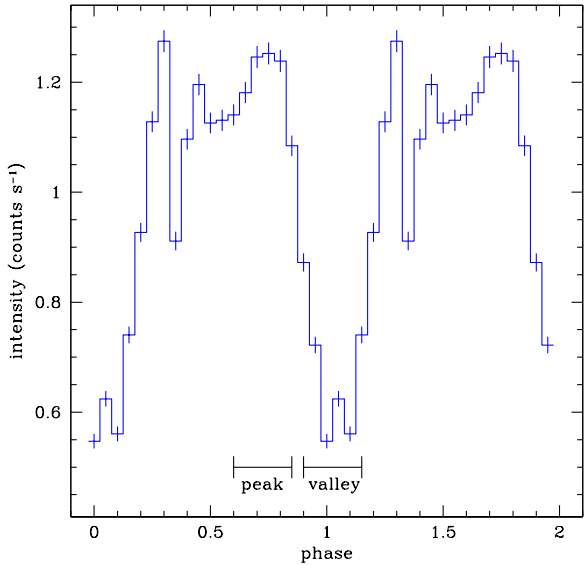}
\caption{Pulse profile showing two phases of the 904-s pulse period in the 3--79 keV energy band beginning at MJD 56834.209667(1). Photons within the intervals labeled ``peak'' and ``valley'' were selected for the phase-resolved spectra. \vspace{1mm}}
\label{fig_fold}
\end{figure}

The known pulsation is detected at a period of 904.0$\pm$0.1 s in the source ($+$ background) light curve with 0.1-s resolution. The best-fitting period is obtained from the fast algorithm for Lomb-Scargle periodograms \citep{lom76,sca82} developed by \citet{pre89}, while the error on the pulse period is derived from the analysis methods of \citet{hor86} and \cite{lea87} which yield consistent uncertainties. Figure \ref{fig_chisq} displays the periodogram and Figure \ref{fig_fold} shows the phase-folded light curve. 

As illustrated in Figure \ref{fig_fold}, the pulse profile begins with a spike in count rates at phase 0.2--0.3, followed by a dip at phase 0.35, and then a broad secondary peak at phase 0.4--0.85. This bimodal pulse profile is similar to the pulse profiles recorded for this source by \integ\ \citep[20--40\,keV:][]{bod06} and \xmm\ \citep[0.3--10\,keV:][]{bod06} as well as with \rxte\ \citep[3--24\,keV:][]{tho06} and \suz\ \citep[0.3--50\,keV:][]{isl15}. 

The pulse fraction ($\equiv \frac{I_{\mathrm{max}} - I_{\mathrm{min}}}{I_{\mathrm{max}} + I_{\mathrm{min}}}$) of 38\%$\pm$1\% is consistent with previous measurements by \integ\ \citep[$57\%\pm24\%$:][]{bod06}, \xmm\ \citep[$38\%\pm5\%$:][]{bod06}, and \suz-XIS \citep[$33\%$:][]{isl15}, but it is higher than the fraction measured with \rxte\ \citep[$21\%\pm1\%$:][]{tho06}. The pulse fraction increases with energy \citep[e.g.,][]{lut09} reaching 60\% for the \nustar\ 10--79-keV energy band, which is consistent with the pulse fraction from \suz-PIN \citep[$65\%$ in 12--50\,keV:][]{isl15}.

\begin{figure}[!t]
\begin{center}
\includegraphics[width=0.45\textwidth]{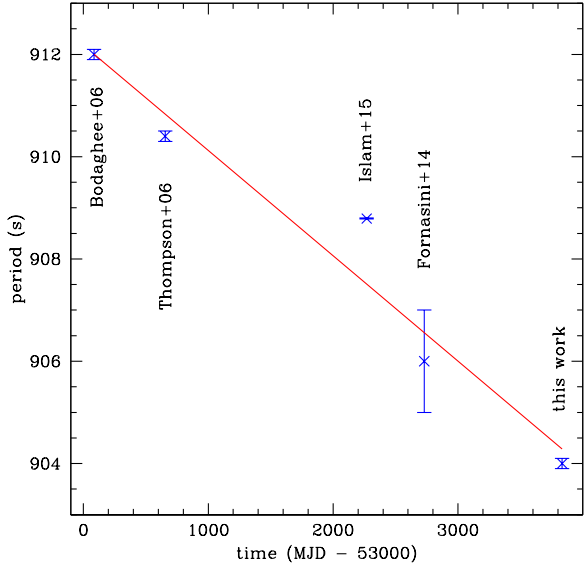}
\end{center}
\caption{Evolution of the pulsation period of IGR~J16393$-$4643 as reported in the literature. A least-squares fit to the data points gives a slope of $\dot P = -2\times10^{-8}$\,s\,s$^{-1}$ (solid line).\vspace{1mm}}
\label{fig_pdot}
\end{figure}

In Figure\,\ref{fig_pdot}, the evolution of the pulsation period is shown as measured by different authors. A least-squares fit to the data results in a slope of $\dot P = -2\times10^{-8}$\,s\,s$^{-1}$, which implies that the neutron star is undergoing a long-term spin-up trend  at a rate of around $-$0.6\,s per year. This is twice the value of $\dot P$ reported by \citet{tho06} using \rxte\ data. The frequency derivative is $\dot \nu = 3\times10^{-14}$ s$^{-2}$. There could be some stochastic variation around the average spin-up value, as seen e.g. in Vela X-1 \citep{tsu89,ikh14}, but the measurements are too sparse to make a definitive claim.

\begin{figure}[!t]
\begin{center}
\includegraphics[width=0.45\textwidth]{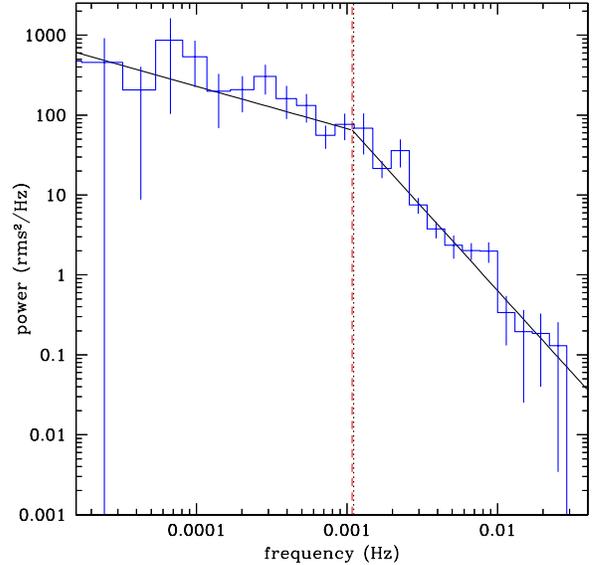}
\end{center}
\caption{Power density spectrum of IGR~J16393$-$4643 after subtraction of the average pulsed periodic component. The expected level of white noise has been subtracted and the spectrum is rms-normalized. The dashed vertical line indicates the pulsation frequency while the dotted vertical line shows the break frequency of a broken power law fit to the data (solid lines).}
\label{fig_psd}
\end{figure}

Figure\,\ref{fig_psd} presents the power density spectrum (PDS) of \igr\ after subtracting the average pulsed periodic component from the light curve. A broken power law model applied to the PDS yields a break frequency of 0.00108(21) Hz with $\Gamma_{1} = -$0.53$\pm$0.13 and $\Gamma_{2} = -$2.08$\pm$0.16. The spectral break is necessary given that its inclusion significantly improves the fit quality ($\chi^{2}/$dof $= 22.8/23$, where dof is the degrees of freedom, compared to $\chi^{2}/$dof $= 94.6/25$ without the break). In contrast to the recent spin-up detection in \object{2RXP~J130159.6$-$635806} \citep{kri15}, where the break in the power spectrum was shifted with respect to the pulsation period, in our case we observe pulsations directly at the break of the power spectrum.

\begin{figure}[!t] 
\centering
\includegraphics[width=0.45\textwidth]{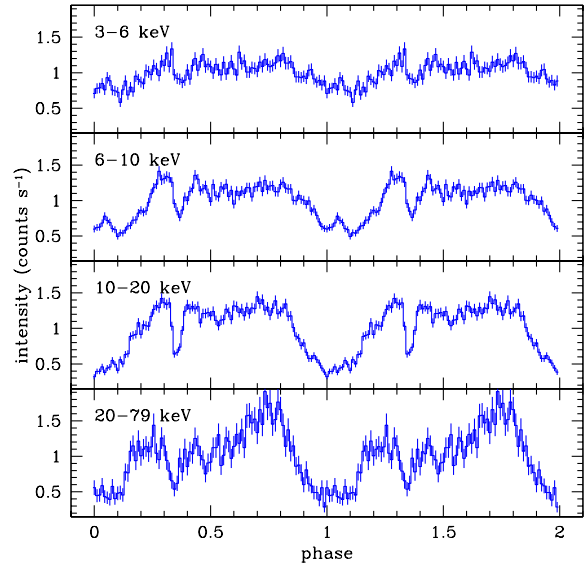}
\caption{Same as Figure \ref{fig_fold} for the 3--6 keV, 6--10 keV, 10--20 keV, and 20--79 keV energy bands with 90 bins per period. In all cases, phase 0 corresponds to MJD 56834.209667(1). }
\label{fig_foldE}
\end{figure}

A phase lag was suggested by \citet{isl15} when they compared the low-energy and high-energy pulse profiles from \suz. The minimum of the pulse profile in the low-energy \nustar\ data (3--10 keV) trails the minimum in the high-energy \nustar\ data (10--79 keV) by 0.1 in phase ($\sim$ 90 s); i.e., consistent with the \suz\ results \citep{isl15}. However, unlike in the \suz\ data, the pulse profiles in the \nustar\ data present two maxima and a dip that are synchronized across energy bands (Figure\,\ref{fig_foldE}).

\subsection{Spectral Analysis}
\label{sec_spec}

The spectral data from \nustar\ FPMA and FPMB were rebinned so that each energy bin had a significance of at least 4$\sigma$. The data were fit simultaneously with an absorbed power law where the instrumental constant was fixed at 1 for FPMA and was allowed to vary for FPMB. In all cases, the instrumental constant that was free to vary remained consistent with 1. The column density was fixed to $2\times10^{23}$ \cmsq\ \citep{bod06}. This yielded a poor fit with a reduced $\chi_{\nu}^{2}/\mathrm{dof} = 7.8/949$ and residuals below 5 keV and above 30 keV. A bremsstrahlung model did not improve the fit by much ($\chi_{\nu}^{2}/\mathrm{dof} = 5.8/949$). 

A thermal blackbody ($kT=4.27$$\pm$0.04 keV) or a power law with an exponential cutoff ($\Gamma = -0.70$$\pm$0.06 and $E_{\mathrm{cut}} = 6.5$$\pm$0.2 keV) offered better fits to the spectral data ($\chi_{\nu}^{2}/\mathrm{dof} = 1.63/949$ and $\chi_{\nu}^{2}/\mathrm{dof} = 1.32/948$, respectively). Some residuals remain near 6.4 keV where an iron K$\alpha$ line is known from \xmm\ observations \citep{bod06}; the addition of a model component for this line leads to a negligible improvement in the fit quality, and so it is not required by the data. The non-detection of this feature is likely due to the lower spectral resolution of \nustar\ at this energy compared with \xmm. Where \nustar\ excels is above 10 keV, and the spectrum shows residuals around 30 keV which suggests a cyclotron resonant scattering feature (CRSF).

To test the significance of this candidate cyclotron line, the \nustar\ spectral data were rebinned so that each energy bin had a significance of least 9$\sigma$. We began with the multi-component continuum model from the broadband spectral fits (see below) with the column density fixed to its best-fitting value (\nh\ $=4\times10^{23}$ \cmsq). Adding a cyclotron line to the model reduced the $\chi^{2}$ of the fit by 30. In order to estimate the significance of the cyclotron line, we relied on the Bayesian posterior predictive probability value (``ppp-value'') as described in \citet{pro02}. See \citet{bel14} and \citet{bha15} for recent applications of this technique. We determined the reference distribution empirically using a Monte Carlo method where we simulated 1000 trials with the \texttt{Xspec} tool \texttt{simftest}, allowing the centroid energy and width of the cyclotron line to vary within their 90\% confidence regions. It is important to note that \texttt{simftest} was used only to simulate the data within a reasonable range of parameter uncertainties and not to perform an F-test. From our simulations, we calculated the change in chi-squared values for the model that includes the cyclotron feature, and for the model without the cyclotron feature (the ``null hypothesis''). These simulations returned a maximum change in $\chi^{2}$ of 14. The probability of finding the observed change in $\chi^{2}$ by chance is $2\times10^{-6}$ which corresponds to 4.8$\sigma$ significance (4.3$\sigma$ after accounting for trials). 

\swift\ spectral data, where each bin contained a minimum of 5 source counts, were then jointly fit with the \nustar\ data. Using single-component models for the continuum led to fits of insufficient quality and unconstrained spectral parameters. The best fit was obtained with a two-component continuum model incorporating both a soft thermal component (a radial blackbody or \texttt{bbodyrad}) and a hard non-thermal component (a power law with an exponential cutoff or \texttt{cutoffpl}). Photoelectric absorption from molecular hydrogen (\texttt{tbabs}) and a cyclotron absorption line (\texttt{cyclabs}) were included \citep[see also][for a more elaborate model]{sch07}. 

\begin{figure}[!t]
\begin{center}
\includegraphics[width=0.45\textwidth,angle=0]{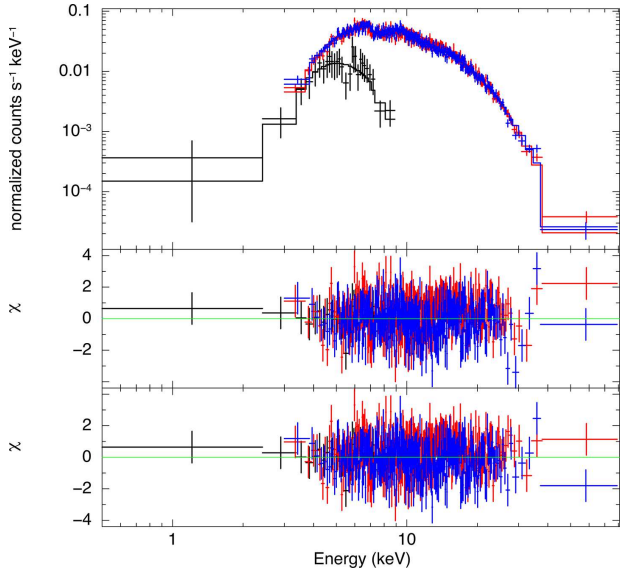}
\end{center}
\caption{Background-subtracted spectra of IGR~J16393$-$4643 collected with \swift\ (black), \nustar-FPMA (blue), and \nustar-FPMB (red). Spectral bins for \swift\ contain a minimum of 5 source counts, while those of \nustar\ have a minimum significance of $7\sigma$. The spectra were fit with a model consisting of an absorbed cutoff power law with a radial blackbody. The middle panel shows residuals from this model fit, while the bottom panel shows residuals when the model includes a cyclotron component. Error bars denote 90\%-confidence limits. The derived spectral parameters (with $\gtrsim 4\sigma$-significance \nustar\ bins) are listed in Table\,\ref{tab_spec}.}
\label{fig_spec}
\end{figure}

The fit quality is excellent as attested by the reduced $\chi_{\nu}^{2}/\mathrm{dof}$ of $0.99/970$ and by the lack of significant residuals as shown in Figure \ref{fig_spec}. Table \ref{tab_spec} lists the spectral parameters of this model. The broadband X-ray flux is about twice as high as was seen previously with \suz\ \citep{isl15} and \xmm\ \citep{bod06}. The effective exposure time is around 51 ks.

To test for variations in the spectral parameters with respect to the pulse profile, we extracted \nustar\ spectra corresponding to the maxima (or ``peak'') and minima (or ``valley'') of the pulse profile. For the peak spectrum, we selected events corresponding to phases 0.6--0.85 (for a count rate of 1.22$\pm$0.01 cps in 3--79 keV), whereas events with phases 0--0.15 and 0.9--1 were chosen for the valley spectrum (a count rate of 0.65$\pm$0.01 cps). For reference, the ``full'' or time-averaged source count rate is 1.05$\pm$0.01 cps.

Each of these phase-resolved \nustar\ spectra covers one-quarter of the full pulse cycle for a total exposure time of $\sim$13 ks per spectrum. Each spectrum was rebinned to have a minimum significance of 6$\sigma$ per bin, and was then jointly fit with the \swift\ spectrum using the multi-component model described above while holding the column density to its optimal value from the time-averaged spectrum (\nh\ $=4.2\times10^{23}$ \cmsq). A few of the spectral parameters could not be easily constrained and so the following results should be considered with that caveat in mind. The cyclotron line energies from both spectra were statistically consistent with each other: $E_{\mathrm{cyc}} =$ 29.4$\pm$1.5 keV for the peak spectrum, and $E_{\mathrm{cyc}} =$ 29.6$\pm$1.6 keV for the valley spectrum. The only significant differences were the normalization and flux values which were twice as high in the peak spectrum, i.e. consistent with the difference in count rates. An instrumental cross-calibration constant near 0.5 for the valley spectrum is expected since the \nustar\ count rate is nearly half its average value while the \swift\ spectrum does not change.

\section{Discussion}
\label{sec_disc}

The X-ray band spectrum generated from joint \swift\ and \nustar\ observations offers the sharpest view yet of the broadband spectral energy distribution (SED) of \igr, especially above $\sim$10 keV. Thermal blackbody photons with a temperature of 1.41$\pm$0.12 keV originating near the surface of the neutron star are re-emitted at higher energies by inverse-Compton scattering off the surrounding electron plasma, and their SED assumes the form a hard power law ($\Gamma = -1.8_{-0.4}^{+0.6}$) with an exponential cutoff at $5.0_{-0.4}^{+0.6}$ keV. Indeed, a thermal Comptonization model also fits the data providing this physical description of the observed phenomena \citep[e.g.,][]{bec05}.

We discovered a cyclotron resonant scattering feature in the \nustar\ spectrum of \igr\ lending further evidence that the compact object in this system is a strongly magnetized neutron star. The centroid energy is $E_{\mathrm{cyc}} = 29.3_{-1.3}^{+1.1}$ keV, and it does not change appreciably across the pulse profile \citep[but see, e.g.,][]{cob02}. This implies a neutron star magnetic field strength of $(2.5\pm0.1)\times10^{12}$ G, which could be up to 40\% larger when gravitational redshift is considered. This $E_{\mathrm{cyc}}$ value is equal to that of \object{X Per} and is within the range of observed line energies in accreting X-ray pulsars \citep[e.g.,][and references therein]{cab12,wal15}. 

The X-ray photons arise at the accretion poles \citep[e.g.,][]{bur91,bec05} given that the emission follows a coherent pulsation period seen in the \nustar\ data of \igr, as well as in previous observations. Since the discovery of the pulse period in 2006, each of the five subsequent observations shows \igr\ spinning at a higher frequency than during the preceding observation. The change in frequency is $\dot \nu = 3\times10^{-14}$\,Hz\,s$^{-1}$ which converts to a long-term spin-up trend of $-$0.6 s per year on a pulsation period of 904.0$\pm$0.1 s in the most recent data (this work). Considering that our period derivative is twice the value found by \citet{tho06}, and considering that the period measurement of \citet{isl15} is not well fit by the least-squares approximation, this indicates that $\dot \nu$ is generally increasing but not at a constant rate.

Angular momentum must be imparted to the neutron star to change the spin period in this way \citep{pri72,lam73}. The environment of the neutron star in an absorbed HMXB such as \igr\ is generally assumed to be a quasi-spherical stellar wind \citep[e.g.,][]{bon52,wal06,sha12,sha15}. Accretion of this wind can, under certain conditions, provide a net change in angular momentum \citep{sha12}. Alternatively, a wind-fed accretion disk can develop around the neutron star in \igr\ \citep{gho79,boe87}, and since a BIII-type donor star is not expected to fill its Roche lobe, this favors a transient scenario for the disk. Transient accretion disks have been suggested in other wind-fed pulsars that exhibit long-term spin-up trends \citep[e.g. \object{GX~301$-$2}:][]{koh97}.

%
\begin{deluxetable}{ l l l l l }
\tabletypesize{\scriptsize}
\tablecolumns{5} 
\tablewidth{0pc} 
\tablecaption{Spectral parameters from the model that best fits the combined \swift-\nustar\ spectrum of IGR~J16393$-$4643. \label{tab_spec}}
\tablehead{
\colhead{model parameter}			& 
\colhead{full}						&  
\colhead{peak}						&  
\colhead{valley}						&      		
\colhead{unit}    			
}
\startdata 

\multicolumn{5}{l}{  \texttt{const$*$tbabs$*$cyclabs$*$(bbodyrad$+$cutoffpl)} \vspace{2mm}} \\

\vspace{2mm}

$C$					& $0.8_{-0.1}^{+0.2}$ 		& $0.9_{-0.1}^{+0.2}$	& 0.5$\pm$0.1	& 				\\

\vspace{2mm}

\nh\					& $42_{-4}^{+3}$ 			& 42			& 42		&	$10^{22}$ \cmsq\		\\

\vspace{2mm}

$\Gamma$			& $-1.8_{-0.4}^{+0.6}$ 		& $\le-2.3$	& $\sim-2.5$	&	 			\\

\vspace{2mm}

$E_{\mathrm{cut}}$		& $5.0_{-0.4}^{+0.6}$ 		& $4.6_{-0.2}^{+0.4}$	& $4.3_{-0.3}^{+0.1}$	&	 keV	\\

\vspace{2mm}

norm. at 1 keV			& 3$\pm$2				& $\sim$1	& $\sim$0.4	&	 $10^{-5}$ ph\,cm$^{-2}$\,s$^{-1}$	\\

\vspace{2mm}

$kT$					& 1.4$\pm$0.1	 			& 1.6$\pm$0.1	& 1.6$\pm$0.1	&	 keV				\\

\vspace{2mm}

norm.				& 0.7$\pm$0.2		 		& $\sim$0.6	& $\sim$0.6	&	 					\\

\vspace{2mm}

$E_{\mathrm{cyc}}$		& $29.3_{-1.3}^{+1.1}$	 	& 29.4$\pm$1.2	& $30.4_{-0.9}^{+1.4}$	&	 keV				\\

\vspace{2mm}

$\sigma_{\mathrm{cyc}}$	& $4_{-2}^{+5}$	 			& $\le5$	& $3_{-1}^{+4}$		&	 keV				\\

\vspace{2mm}

depth				& 0.4$\pm$0.1	 			& $0.6_{-0.3}^{+0.4}$	& $1.9_{-0.6}^{+1.6}$		&	 keV				\\

\vspace{2mm}

$F_{\mathrm{abs}}$		& $19.6_{-1.4}^{+0.3}$		& 22$\pm$11	& $12_{-7}^{+4}$ 	&	 $10^{-11}$ \ergcms\	\\

\vspace{2mm}

$F_{\mathrm{unabs}}$	& $28.3_{-2.0}^{+1.3}$		& 31$\pm$15	& 18$\pm$8	&	 $10^{-11}$ \ergcms\	\\

\vspace{2mm}

$\chi_{\nu}^{2}/\mathrm{dof}$	& $0.99/970$				& $1.06/749$	& $1.10/749$	&	 	

\enddata
\tablecomments{Spectral parameters are shown for the ``full'' or time-averaged spectrum, as well as for the 904-s pulse phase-resolved spectra where the ``peak'' corresponds to counts from phases 0.6--0.85, while the ``valley'' includes only those counts corresponding to phases in 0.0--1.5 and 0.9--1.0 (see Figure \ref{fig_fold}). The column density for the phase-resolved spectra is fixed to the value from the ``full'' spectrum. $C$ is an instrumental cross-calibration coefficient which is fixed at 1 for \swift\ and variable for \nustar. Flux values are given as observed (``abs'') and corrected for absorption (``unabs'') in the 0.5--80-keV energy range. Uncertainties on the flux offer a more realistic representation of the error range than do the uncertainties on the normalizations which are large and mostly omitted. Errors are quoted at 90\% confidence. }
\end{deluxetable}

Naturally, the strong magnetic field of the pulsar influences the dynamics of the accretion disk out to a certain distance \citep{gho79}. Material within the corotation radius is forced to corotate with the neutron star and if its angular velocity is less than the local Keplerian velocity, then the material can be accreted along the field lines (this implies a magnetospheric radius smaller than the corotation radius). Depending on whether the accreted material's angular momentum has the same or an opposite direction as the neutron star's spin, this will cause the neutron star to spin faster or slower, respectively \citep{wat89}. The consistent downward trend of the pulsation period in \igr\ suggests that material in the accretion disk orbits in the same direction as the neutron star's spin. 

Additional accretion dynamics can be elucidated from the power density spectrum (PDS). Going from low frequency to high frequency, the PDS changes from a flat power law (slope $\sim -0.5$) to a steeper power law (slope $\sim -2$). The PDS breaks at a frequency of 0.00108(21) Hz which is very close to the spin frequency of the neutron star (0.0011062(1) Hz). These characteristics of the PDS in persistent, accreting X-ray pulsars are described by \citet{rev09}, and they suggest that the neutron star in \igr\ is spinning close to corotation with the inner edge of an accretion disk that has been truncated by the magnetic field. Using canonical values for the neutron star mass ($1.4 M_{\odot}$) and magnetic dipole moment ($\mu = 10^{30}$\,G\,cm$^{-3}$ for $B \sim 10^{12}$\,G), this gives estimates of $1.1\times10^{8}$\,cm and $2.0\times10^{8}$\,cm for the magnetospheric radius assuming mass accretion rates of $10^{-6}$\,$M_{\odot}$\,yr$^{-1}$ and $10^{-7}$\,$M_{\odot}$\,yr$^{-1}$, respectively. However, an alternative model for quasi-spherical wind accretion developed by \citet{sha12} could also produce the observed change in spin frequency, and in this case, an accretion disk is not required.

\section{Summary \& Conclusions}
\label{sec_conc}

A \nustar\ observation of \igr\ has revealed valuable insights into the nature of this source. The detection of a cyclotron resonant scattering feature at $29.3_{-1.3}^{+1.1}$ keV allowed us to constrain the magnetic field strength to $B = (2.5\pm0.1)\times10^{12}$ G. This is the first time that the magnetic field has been measured in this object. The cyclotron line was not detected in previous excursions into this energy range by \rxte-ASM, \integ-ISGRI, and \suz-HXD. This result is a testament to the spectral sensitivity and resolving power of \nustar. 

The pulsation period of the neutron star in \igr\ is now at 904.0$\pm$0.1\,s. Looking at the five measurements made between 2006 and 2014, we find that this period has consistently gotten shorter with time at an average long-term spin-up rate of $-$0.6\,s per year ($\dot \nu = 3\times10^{-14}$ Hz\,s$^{-1}$). The slope of the power density spectrum breaks near the pulsation frequency as expected for persistently emitting accretion-powered pulsars. This could indicate that a transient and magnetically truncated accretion disk is almost in corotation with the neutron star whose magnetospheric radius is around $2\times10^{8}$\,cm, although accretion from a quasi-spherical wind could also lead to the observed change in pulsation frequency.

\igr\ is an archetype of the class of heavily obscured wind-accreting pulsars, with its ks-long pulsation period and its column density that is at least an order of magnitude greater than the expected line-of-sight value. Yet its source classification has been subject to multiple changes over the years and the identity of the donor star remains elusive. Results we obtained thanks to \nustar\ represent new puzzle pieces to add to the still-developing picture we have of this intriguing source.

\acknowledgments
The authors are grateful to the anonymous Referee for their constructive criticism that helped improve the quality of the manuscript. AB thanks Dr. Konstantin Postnov. FMF acknowledges support from the National Science Foundation Graduate Research Fellowship. RK acknowledges support from from Russian Science Foundation (grant 14-22-00271). The scientific results reported in this article are based on data from the \nustar\ mission, a project led by the California Institute of Technology, managed by the Jet Propulsion Laboratory, and funded by the National Aeronautics and Space Administration. We thank the \nustar\ Operations, Software, and Calibration teams for support with the execution and analysis of these observations. This research has made use of: the \nustar\ Data Analysis Software (NuSTARDAS) jointly developed by the ASI Science Data Center (ASDC, Italy) and the California Institute of Technology; data obtained from the High Energy Astrophysics Science Archive Research Center (HEASARC) provided by NASA's Goddard Space Flight Center; NASA's Astrophysics Data System Bibliographic Services; and the SIMBAD database operated at CDS, Strasbourg, France.

{\it Facilities:} \facility{NuSTAR}, \facility{Swift}

\bibliographystyle{apj}
\bibliography{bod.bib}

\begin{thebibliography}{}
\expandafter\ifx\csname natexlab\endcsname\relax\def\natexlab#1{#1}\fi

\bibitem[{{Arnaud}(1996)}]{arn96}
{Arnaud}, K.~A. 1996, in Astronomical Society of the Pacific Conference Series,
  Vol. 101, Astronomical Data Analysis Software and Systems V, ed. G.~H.
  {Jacoby} \& J.~{Barnes}, 17

\bibitem[{{Becker} \& {Wolff}(2005)}]{bec05}
{Becker}, P.~A., \& {Wolff}, M.~T. 2005, \apj, 630, 465

\bibitem[{{Bellm} {et~al.}(2014){Bellm}, {F{\"u}rst}, {Pottschmidt}, {Tomsick},
  {Boggs}, {Chakrabarty}, {Christensen}, {Craig}, {Hailey}, {Harrison},
  {Stern}, {Walton}, {Wilms}, \& {Zhang}}]{bel14}
{Bellm}, E.~C., {F{\"u}rst}, F., {Pottschmidt}, K., {et~al.} 2014, \apj, 792,
  108

\bibitem[{{Bhalerao} {et~al.}(2015){Bhalerao}, {Romano}, {Tomsick},
  {Natalucci}, {Smith}, {Bellm}, {Boggs}, {Chakrabarty}, {Christensen},
  {Craig}, {Fuerst}, {Hailey}, {Harrison}, {Krivonos}, {Lu}, {Madsen}, {Stern},
  {Younes}, \& {Zhang}}]{bha15}
{Bhalerao}, V., {Romano}, P., {Tomsick}, J., {et~al.} 2015, \mnras, 447, 2274

\bibitem[{{Bildsten} {et~al.}(1997){Bildsten}, {Chakrabarty}, {Chiu}, {Finger},
  {Koh}, {Nelson}, {Prince}, {Rubin}, {Scott}, {Stollberg}, {Vaughan},
  {Wilson}, \& {Wilson}}]{bil97}
{Bildsten}, L., {Chakrabarty}, D., {Chiu}, J., {et~al.} 1997, \apjs, 113, 367

\bibitem[{{Bird} {et~al.}(2004){Bird}, {Barlow}, {Bassani}, {Bazzano},
  {Bodaghee}, {Capitanio}, {Cocchi}, {Del Santo}, {Dean}, {Hill}, {Lebrun},
  {Malaguti}, {Malizia}, {Much}, {Shaw}, {Stephen}, {Terrier}, {Ubertini}, \&
  {Walter}}]{bir04}
{Bird}, A.~J., {Barlow}, E.~J., {Bassani}, L., {et~al.} 2004, \apjl, 607, L33

\bibitem[{{Bodaghee} {et~al.}(2012){Bodaghee}, {Rahoui}, {Tomsick}, \&
  {Rodriguez}}]{bod12b}
{Bodaghee}, A., {Rahoui}, F., {Tomsick}, J.~A., \& {Rodriguez}, J. 2012, \apj,
  751, 113

\bibitem[{{Bodaghee} {et~al.}(2006){Bodaghee}, {Walter}, {Zurita Heras},
  {Bird}, {Courvoisier}, {Malizia}, {Terrier}, \& {Ubertini}}]{bod06}
{Bodaghee}, A., {Walter}, R., {Zurita Heras}, J.~A., {et~al.} 2006, \aap, 447,
  1027

\bibitem[{{Bodaghee} {et~al.}(2014){Bodaghee}, {Tomsick}, {Krivonos}, {Stern},
  {Bauer}, {Fornasini}, {Barri{\`e}re}, {Boggs}, {Christensen}, {Craig},
  {Gotthelf}, {Hailey}, {Harrison}, {Hong}, {Mori}, \& {Zhang}}]{bod14}
{Bodaghee}, A., {Tomsick}, J.~A., {Krivonos}, R., {et~al.} 2014, \apj, 791, 68

\bibitem[{{Boerner} {et~al.}(1987){Boerner}, {Hayakawa}, {Nagase}, \&
  {Anzer}}]{boe87}
{Boerner}, G., {Hayakawa}, S., {Nagase}, F., \& {Anzer}, U. 1987, \aap, 182, 63

\bibitem[{{Bondi}(1952)}]{bon52}
{Bondi}, H. 1952, \mnras, 112, 195

\bibitem[{{Burnard} {et~al.}(1991){Burnard}, {Arons}, \& {Klein}}]{bur91}
{Burnard}, D.~J., {Arons}, J., \& {Klein}, R.~I. 1991, \apj, 367, 575

\bibitem[{{Burrows} {et~al.}(2005){Burrows}, {Hill}, {Nousek}, {Kennea},
  {Wells}, {Osborne}, {Abbey}, {Beardmore}, {Mukerjee}, {Short}, {Chincarini},
  {Campana}, {Citterio}, {Moretti}, {Pagani}, {Tagliaferri}, {Giommi},
  {Capalbi}, {Tamburelli}, {Angelini}, {Cusumano}, {Br{\"a}uninger}, {Burkert},
  \& {Hartner}}]{bur05}
{Burrows}, D.~N., {Hill}, J.~E., {Nousek}, J.~A., {et~al.} 2005, \ssr, 120, 165

\bibitem[{{Caballero} \& {Wilms}(2012)}]{cab12}
{Caballero}, I., \& {Wilms}, J. 2012, \memsai, 83, 230

\bibitem[{{Chaty} {et~al.}(2008){Chaty}, {Rahoui}, {Foellmi}, {Tomsick},
  {Rodriguez}, \& {Walter}}]{cha08}
{Chaty}, S., {Rahoui}, F., {Foellmi}, C., {et~al.} 2008, \aap, 484, 783

\bibitem[{{Coburn} {et~al.}(2002){Coburn}, {Heindl}, {Rothschild}, {Gruber},
  {Kreykenbohm}, {Wilms}, {Kretschmar}, \& {Staubert}}]{cob02}
{Coburn}, W., {Heindl}, W.~A., {Rothschild}, R.~E., {et~al.} 2002, \apj, 580,
  394

\bibitem[{{Coley} {et~al.}(2015){Coley}, {Corbet}, \& {Krimm}}]{col15}
{Coley}, J.~B., {Corbet}, R.~H.~D., \& {Krimm}, H.~A. 2015, \apj, 808, 140

\bibitem[{{Combi} {et~al.}(2004){Combi}, {Rib{\'o}}, {Mirabel}, \&
  {Sugizaki}}]{com04}
{Combi}, J.~A., {Rib{\'o}}, M., {Mirabel}, I.~F., \& {Sugizaki}, M. 2004, \aap,
  422, 1031

\bibitem[{{Corbet}(1986)}]{cor86}
{Corbet}, R.~H.~D. 1986, \mnras, 220, 1047

\bibitem[{{Corbet} \& {Krimm}(2013)}]{cor13}
{Corbet}, R.~H.~D., \& {Krimm}, H.~A. 2013, \apj, 778, 45

\bibitem[{{Corbet} {et~al.}(2010){Corbet}, {Krimm}, {Barthelmy}, {Baumgartner},
  {Markwardt}, {Skinner}, \& {Tueller}}]{cor10}
{Corbet}, R.~H.~D., {Krimm}, H.~A., {Barthelmy}, S.~D., {et~al.} 2010, The
  Astronomer's Telegram, 2570, 1

\bibitem[{{Cutri} {et~al.}(2003){Cutri}, {Skrutskie}, {van Dyk}, {Beichman},
  {Carpenter}, {Chester}, {Cambresy}, {Evans}, {Fowler}, {Gizis}, {Howard},
  {Huchra}, {Jarrett}, {Kopan}, {Kirkpatrick}, {Light}, {Marsh}, {McCallon},
  {Schneider}, {Stiening}, {Sykes}, {Weinberg}, {Wheaton}, {Wheelock}, \&
  {Zacarias}}]{cut03}
{Cutri}, R.~M., {Skrutskie}, M.~F., {van Dyk}, S., {et~al.} 2003, {2MASS All
  Sky Catalog of point sources.} ({Cutri, R.~M., Skrutskie, M.~F., van Dyk, S.,
  Beichman, C.~A., Carpenter, J.~M., Chester, T., Cambresy, L., Evans, T.,
  Fowler, J., Gizis, J., Howard, E., Huchra, J., Jarrett, T., Kopan, E.~L.,
  Kirkpatrick, J.~D., Light, R.~M., Marsh, K.~A., McCallon, H., Schneider, S.,
  Stiening, R., Sykes, M., Weinberg, M., Wheaton, W.~A., Wheelock, S., \&
  Zacarias, N.})

\bibitem[{{Fornasini} {et~al.}(2014){Fornasini}, {Tomsick}, {Bodaghee},
  {Krivonos}, {An}, {Rahoui}, {Gotthelf}, {Bauer}, \& {Stern}}]{for14}
{Fornasini}, F.~M., {Tomsick}, J.~A., {Bodaghee}, A., {et~al.} 2014, \apj, 796,
  105

\bibitem[{{Ghosh} \& {Lamb}(1979)}]{gho79}
{Ghosh}, P., \& {Lamb}, F.~K. 1979, \apj, 234, 296

\bibitem[{{Harrison} {et~al.}(2013){Harrison}, {Craig}, {Christensen},
  {Hailey}, {Zhang}, {Boggs}, {Stern}, {Cook}, {Forster}, {Giommi},
  {Grefenstette}, {Kim}, {Kitaguchi}, {Koglin}, {Madsen}, {Mao}, {Miyasaka},
  {Mori}, {Perri}, {Pivovaroff}, {Puccetti}, {Rana}, {Westergaard}, {Willis},
  {Zoglauer}, {An}, {Bachetti}, {Barri{\`e}re}, {Bellm}, {Bhalerao},
  {Brejnholt}, {Fuerst}, {Liebe}, {Markwardt}, {Nynka}, {Vogel}, {Walton},
  {Wik}, {Alexander}, {Cominsky}, {Hornschemeier}, {Hornstrup}, {Kaspi},
  {Madejski}, {Matt}, {Molendi}, {Smith}, {Tomsick}, {Ajello}, {Ballantyne},
  {Balokovi{\'c}}, {Barret}, {Bauer}, {Blandford}, {Brandt}, {Brenneman},
  {Chiang}, {Chakrabarty}, {Chenevez}, {Comastri}, {Dufour}, {Elvis}, {Fabian},
  {Farrah}, {Fryer}, {Gotthelf}, {Grindlay}, {Helfand}, {Krivonos}, {Meier},
  {Miller}, {Natalucci}, {Ogle}, {Ofek}, {Ptak}, {Reynolds}, {Rigby},
  {Tagliaferri}, {Thorsett}, {Treister}, \& {Urry}}]{har13}
{Harrison}, F.~A., {Craig}, W.~W., {Christensen}, F.~E., {et~al.} 2013, \apj,
  770, 103

\bibitem[{{Hartman} {et~al.}(1999){Hartman}, {Bertsch}, {Bloom}, {Chen},
  {Deines-Jones}, {Esposito}, {Fichtel}, {Friedlander}, {Hunter}, {McDonald},
  {Sreekumar}, {Thompson}, {Jones}, {Lin}, {Michelson}, {Nolan}, {Tompkins},
  {Kanbach}, {Mayer-Hasselwander}, {M{\"u}cke}, {Pohl}, {Reimer}, {Kniffen},
  {Schneid}, {von Montigny}, {Mukherjee}, \& {Dingus}}]{har99}
{Hartman}, R.~C., {Bertsch}, D.~L., {Bloom}, S.~D., {et~al.} 1999, \apjs, 123,
  79

\bibitem[{{Horne} \& {Baliunas}(1986)}]{hor86}
{Horne}, J.~H., \& {Baliunas}, S.~L. 1986, \apj, 302, 757

\bibitem[{{Ikhsanov} {et~al.}(2014){Ikhsanov}, {Likh}, \&
  {Beskrovnaya}}]{ikh14}
{Ikhsanov}, N.~R., {Likh}, Y.~S., \& {Beskrovnaya}, N.~G. 2014, Astronomy
  Reports, 58, 376

\bibitem[{{Islam} {et~al.}(2015){Islam}, {Maitra}, {Pradhan}, \&
  {Paul}}]{isl15}
{Islam}, N., {Maitra}, C., {Pradhan}, P., \& {Paul}, B. 2015, \mnras, 446, 4148

\bibitem[{{Koh} {et~al.}(1997){Koh}, {Bildsten}, {Chakrabarty}, {Nelson},
  {Prince}, {Vaughan}, {Finger}, {Wilson}, \& {Rubin}}]{koh97}
{Koh}, D.~T., {Bildsten}, L., {Chakrabarty}, D., {et~al.} 1997, \apj, 479, 933

\bibitem[{{Krivonos} {et~al.}(2015){Krivonos}, {Tsygankov}, {Lutovinov},
  {Tomsick}, {Chakrabarty}, {Bachetti}, {Boggs}, {Chernyakova}, {Christensen},
  {Craig}, {F{\"u}rst}, {Hailey}, {Harrison}, {Lansbury}, {Rahoui}, {Stern}, \&
  {Zhang}}]{kri15}
{Krivonos}, R.~A., {Tsygankov}, S.~S., {Lutovinov}, A.~A., {et~al.} 2015, \apj,
  809, 140

\bibitem[{{Lamb} {et~al.}(1973){Lamb}, {Pethick}, \& {Pines}}]{lam73}
{Lamb}, F.~K., {Pethick}, C.~J., \& {Pines}, D. 1973, \apj, 184, 271

\bibitem[{{Leahy}(1987)}]{lea87}
{Leahy}, D.~A. 1987, \aap, 180, 275

\bibitem[{{Lomb}(1976)}]{lom76}
{Lomb}, N.~R. 1976, \apss, 39, 447

\bibitem[{{Lutovinov} \& {Tsygankov}(2009)}]{lut09}
{Lutovinov}, A.~A., \& {Tsygankov}, S.~S. 2009, Astronomy Letters, 35, 433

\bibitem[{{Madsen} {et~al.}(2015){Madsen}, {Harrison}, {Markwardt}, {An},
  {Grefenstette}, {Bachetti}, {Miyasaka}, {Kitaguchi}, {Bhalerao},
  {Christensen}, {Craig}, {Fuerst}, {Walton}, {Hailey}, {Rana}, {Stern},
  {Westergaard}, \& {Zhang}}]{mad15}
{Madsen}, K.~K., {Harrison}, F.~A., {Markwardt}, C., {et~al.} 2015, ArXiv
  e-prints: 1504.01672, arXiv:1504.01672

\bibitem[{{Malizia} {et~al.}(2004){Malizia}, {Bassani}, {Di Cocco}, {Stephen},
  {Walter}, {Bodaghee}, \& {Bazzano}}]{mal04}
{Malizia}, A., {Bassani}, L., {Di Cocco}, G., {et~al.} 2004, The Astronomer's
  Telegram, 227, 1

\bibitem[{{Nagase}(1989)}]{nag89}
{Nagase}, F. 1989, \pasj, 41, 1

\bibitem[{{Nespoli} {et~al.}(2010){Nespoli}, {Fabregat}, \&
  {Mennickent}}]{nes10}
{Nespoli}, E., {Fabregat}, J., \& {Mennickent}, R.~E. 2010, \aap, 516, A94

\bibitem[{{Press} \& {Rybicki}(1989)}]{pre89}
{Press}, W.~H., \& {Rybicki}, G.~B. 1989, \apj, 338, 277

\bibitem[{{Pringle} \& {Rees}(1972)}]{pri72}
{Pringle}, J.~E., \& {Rees}, M.~J. 1972, \aap, 21, 1

\bibitem[{{Protassov} {et~al.}(2002){Protassov}, {van Dyk}, {Connors},
  {Kashyap}, \& {Siemiginowska}}]{pro02}
{Protassov}, R., {van Dyk}, D.~A., {Connors}, A., {Kashyap}, V.~L., \&
  {Siemiginowska}, A. 2002, \apj, 571, 545

\bibitem[{{Revnivtsev} {et~al.}(2009){Revnivtsev}, {Churazov}, {Postnov}, \&
  {Tsygankov}}]{rev09}
{Revnivtsev}, M., {Churazov}, E., {Postnov}, K., \& {Tsygankov}, S. 2009, \aap,
  507, 1211

\bibitem[{{Scargle}(1982)}]{sca82}
{Scargle}, J.~D. 1982, \apj, 263, 835

\bibitem[{{Sch{\"o}nherr} {et~al.}(2007){Sch{\"o}nherr}, {Wilms}, {Kretschmar},
  {Kreykenbohm}, {Santangelo}, {Rothschild}, {Coburn}, \& {Staubert}}]{sch07}
{Sch{\"o}nherr}, G., {Wilms}, J., {Kretschmar}, P., {et~al.} 2007, \aap, 472,
  353

\bibitem[{{Shakura} {et~al.}(2012){Shakura}, {Postnov}, {Kochetkova}, \&
  {Hjalmarsdotter}}]{sha12}
{Shakura}, N., {Postnov}, K., {Kochetkova}, A., \& {Hjalmarsdotter}, L. 2012,
  \mnras, 420, 216

\bibitem[{{Shakura} {et~al.}(2015){Shakura}, {Postnov}, {Kochetkova},
  {Hjalmarsdotter}, {Sidoli}, \& {Paizis}}]{sha15}
{Shakura}, N.~I., {Postnov}, K.~A., {Kochetkova}, A.~Y., {et~al.} 2015,
  Astronomy Reports, 59, 645

\bibitem[{{Sugizaki} {et~al.}(2001){Sugizaki}, {Mitsuda}, {Kaneda},
  {Matsuzaki}, {Yamauchi}, \& {Koyama}}]{sug01}
{Sugizaki}, M., {Mitsuda}, K., {Kaneda}, H., {et~al.} 2001, \apjs, 134, 77

\bibitem[{{Thompson} {et~al.}(2006){Thompson}, {Tomsick}, {Rothschild}, {in't
  Zand}, \& {Walter}}]{tho06}
{Thompson}, T.~W.~J., {Tomsick}, J.~A., {Rothschild}, R.~E., {in't Zand},
  J.~J.~M., \& {Walter}, R. 2006, \apj, 649, 373

\bibitem[{{Tsunemi}(1989)}]{tsu89}
{Tsunemi}, H. 1989, \pasj, 41, 453

\bibitem[{{Verner} {et~al.}(1996){Verner}, {Ferland}, {Korista}, \&
  {Yakovlev}}]{ver96}
{Verner}, D.~A., {Ferland}, G.~J., {Korista}, K.~T., \& {Yakovlev}, D.~G. 1996,
  \apj, 465, 487

\bibitem[{{Walter} {et~al.}(2015){Walter}, {Lutovinov}, {Bozzo}, \&
  {Tsygankov}}]{wal15}
{Walter}, R., {Lutovinov}, A.~A., {Bozzo}, E., \& {Tsygankov}, S.~S. 2015,
  \aapr, 23, 2

\bibitem[{{Walter} {et~al.}(2006){Walter}, {Zurita Heras}, {Bassani},
  {Bazzano}, {Bodaghee}, {Dean}, {Dubath}, {Parmar}, {Renaud}, \&
  {Ubertini}}]{wal06}
{Walter}, R., {Zurita Heras}, J., {Bassani}, L., {et~al.} 2006, \aap, 453, 133

\bibitem[{{Waters} \& {van Kerkwijk}(1989)}]{wat89}
{Waters}, L.~B.~F.~M., \& {van Kerkwijk}, M.~H. 1989, \aap, 223, 196

\bibitem[{{White} {et~al.}(1983){White}, {Swank}, \& {Holt}}]{whi83}
{White}, N.~E., {Swank}, J.~H., \& {Holt}, S.~S. 1983, \apj, 270, 711

\bibitem[{{Wilms} {et~al.}(2000){Wilms}, {Allen}, \& {McCray}}]{wil00}
{Wilms}, J., {Allen}, A., \& {McCray}, R. 2000, \apj, 542, 914

\end{thebibliography}

\clearpage

\end{document}